\documentclass[showpacs,preprintnumbers,amsmath,amssymb,12pt,floatfix,epsfig]{revtex4}
\usepackage{dcolumn}
\usepackage{bm}
\usepackage[dvips]{color}
\usepackage{graphicx}

\baselineskip=24ptplus.5ptminus.2pt \vspace*{0.1 in} \large

\begin{document}
\title{\bf Coulomb sum rule in quasielastic region}
\author{K. S. Kim\footnote{E-mail : kyungsik@hau.ac.kr} and B. G. Yu}
\affiliation{\it School of Liberal Arts and Science, Hankuk
Aviation University, Koyang 412-791, Korea}
\author{M. K. Cheoun}
\affiliation{\it Department of Physics, Soongsil University,
Seoul, 156-743, Korea}

\begin{abstract}
Within a relativistic single particle model, we calculate the
Coulomb sum rule of inclusive electron scattering from $^{40}$Ca
and $^{208}$Pb in quasielastic region. Theoretical longitudinal
and transverse structure functions are extracted for three
momentum transfers from 300 to 500 MeV/c and compared with the
experimental data measured at Bates and Saclay. We find that there
is no drastic suppression of the longitudinal structure function
and that the Coulomb sum rule depends on nucleus in our
theoretical model.
\end{abstract}

\maketitle

Medium energy electron scattering in quasielastic region has been
acknowledged one of most useful tools to investigate not only the
nuclear structure and properties but also the nucleon properties
in nuclear medium. It is of considerable interest in experimental
and theoretical aspects to extract longitudinal and transverse
structure functions as a function of energy transfer at fixed
three momentum transfer, because both structure functions stand
for the electric and magnetic responses of the target nucleus,
respectively.

The Fermi gas model in the impulse approximation describes roughly
the inclusive $(e,e')$ cross sections but fails to reproduce the
structure functions. In particular, there appeared to be a large
suppression (about 50\%) of the longitudinal structure function,
referred to the missing strength in the Coulomb sum rule (CSR)
\cite{mezi}, which states that the integration of the (charge
response) longitudinal function over the full range of energy loss
should be equal to the charge of the nucleus. Many results of the
CSR for several nuclei such as $^{12}$C \cite{barr}, $^{40}$Ca
\cite{mezi,dead,bates}, and $^{208}$Pb \cite{sacl} have been
reaped during the last two decades.

However the CSR could be affected by the nucleons' correlations,
such as Pauli, long-range and short-range correlations, which are
inescapable drawbacks in the mean field approach to the many
nucleon systems\cite{bates}. For example, Pauli correlations
simply due to the fermion nature of the nucleon cause some
deviations of the CSR from the proton number $Z$ at the low
momentum transfer region \cite{jourdan}. Fabrocini and Fantoni
\cite{fabr}, for instance, found that the CSR within the framework
of a non-relativistic nuclear matter calculation was saturated
much slower than the calculation by the Fermi gas model. The
short-range correlations between nucleons may also lead to the
deviations of the CSR from the proton number $Z$. But such
correlations due to the bounded nucleons turned out to account for
at most 10 \% reduction from the $Z$ number \cite{workshop05}.

Therefore any further suppression would indicate some
modifications of the nucleon by the nuclear medium. This causes
the compelling and revivable motivation to investigate the Coulomb
sum rule in both theoretical and experimental sides \cite {orla}.
Of course, the prerequisite before meaningful debates about the
medium effect is whether the suppression exist on the CSR, but the
missing strength is still an unsolved problem in spite of lots of
theoretical efforts.

The CSR on $^{40}$Ca measured from Bates \cite{dead,bates} showed
a suppression of about $30$\% in the $q$ range from 300 MeV/c to
450 MeV/c. But the CSR on $^{208}$Pb at Saclay \cite{sacl} showed
a reduction of about 50\% in the effective momentum transfer from
350 MeV/c to 550 MeV/c. What these measurements depend on the
nucleus might not be clearly understood. While Traini \cite{trai}
insisted on the suppression of the CSR which depends on the
nucleus, Jourdan \cite{jourdan} concluded that the longitudinal
structure function is not suppressed and there is no A-dependent
quenching for the Coulomb sum in the analysis of the experimental
data for $^{12}$C, $^{40}$Ca, and $^{56}$Fe.

Morgenstern and Meziani \cite{morg}, on the contrary, reanalyzed
the experimental data of $^{40}$Ca, $^{48}$Ca, $^{56}$Fe,
$^{197}$Au, $^{208}$Pb, and $^{238}$U at the effective momentum
transfer 500 MeV/c. They claimed that the suppression of the
longitudinal structure function exists about 40 \% at the
effective momentum transfer 500 MeV/c, and tried to explain the
suppression for a change of the nucleon properties inside the
nuclear medium.

Under these unsettled and controversial situations, future
progress on the CSR would be made only by the expected JLAB
results at the high momentum transfer region, in which the
relevant correlations become small. But in that region, the
Coulomb corrections between the electron and the nucleus have to
be pinned down to extract the information of the structure
functions to understand the CSR. As detailed in Ref.
\cite{workshop05,kim05}, there are two different methods,
effective momentum approximation (EMA) and full calculations just
like Ohio group's calculations. Although at high momentum transfer
region nearly same results are expected from both approaches
\cite{kim05,and05}, there still remained some controversy below
the momentum region, so that one keenly needs to compare both
results at the region.

As discussed in our previous papers \cite{kim96,kim01,kim03}, our
approximate treatment of the electron Coulomb distortion from
medium and heavy nuclei turned out to agree to a full distorted
wave Born approximation (DWBA) within a few percent (see Ref.
\cite{kim96,kim01} in detail). This approximation allows the
separation of the cross section into a longitudinal term and a
transverse term while the full DWBA calculation cannot yield the
separation. In this work, this approximation is taken into account
in all the calculations in order to include the electron Coulomb
distortion.

In this work, therefore, along the Ohio model we investigate the
CSR by comparing with Bates and Saclay data in the $q$ range from
300 MeV/c to 500 MeV/c in quasielastic region, whose recent
results and detailed discussions appeared Ref. \cite{kim05}. In
order to calculate the nuclear transition current, we use the
relativistic single particle model for the bound state wave
function in the presence of the strong scalar and vector
potentials based on the $\sigma - \omega$ model generated by
Horowitz and Serot \cite{horo}.

In the plane wave Born approximation (PWBA) in which the electrons
are described as Dirac plane waves, the cross section for the
inclusive $(e,e')$ scattering can be written as
\begin{equation}
{\frac {d^2 \sigma} {d\Omega d\omega} } = \sigma_M  \left [ {\frac
{q^4_{\mu}} {q^4}} R_L (q,\omega) + (\tan^2 {\frac {\theta} {2}} -
{\frac {q^2_{\mu}} {2q^2}} )R_T (q,\omega) \right ], \label{csr}
\end{equation}
where $q^2_{\mu}=\omega^2 - {\bf q}^2 = -Q^2$ is the four momentum
transfer, $\sigma_M$ is the Mott cross section, and $R_L$ and
$R_T$ are the longitudinal and transverse structure functions
which depend only on the three momentum transfer $q$ and the
energy transfer $\omega$. Note that the scalar and vector
potentials for the outgoing nucleons are used as the
energy-dependent form in Ref. \cite{kim03}. These different
potentials between the bound and continuum states result in
non-conserved current. Hence instead of eliminating the
$z$-component we directly calculate the $z$-component of the
nucleon current. Notice that this energy-dependent potential still
includes the final state interaction for the outgoing nucleon.

From the measured cross section in Eq. (\ref{csr}), the total
structure function is defined as
\begin{equation}
S_{tot} (q, \omega, \theta) = \left ( {\frac {\epsilon (\theta)}
{\sigma_M}} \right) \left ({\frac {q^4}{Q^4}} \right) {\frac {d^2
\sigma} {d\Omega d\omega} }, \label{stotal}
\end{equation}
where the $\epsilon ( \theta )$ is the virtual photon
polarization.

Therefore, the total structure function in Eq. (\ref{stotal})
becomes
\begin{equation}
S_{tot} (q, \omega, \theta) = \epsilon (\theta) R_L (q, \omega) +
\left ( {\frac {q^2} {2 Q^2}} \right ) R_T (q,\omega).
\label{stot}
\end{equation}
$S_{tot}$ is described as a straight line in terms of the
independent variable $\epsilon(\theta)$ with slope $R_L(q,
\omega)$ and intercept proportional to $R_T(q, \omega)$ by keeping
the momentum transfer $q$ and the energy transfer $\omega$ fixed.

The CSR is defined as the integration of the total longitudinal
structure function in Eq. (\ref{stot}) for inclusive $(e,e')$
reaction by de Forest \cite{defo2}
\begin{equation}
C(q) = {\frac {1} {Z}} \int^{\infty}_{\omega_{min}} {\frac
{R_{L}(q,\omega)} {\tilde{G}_{E}^{2} (Q^2)}} d\omega, \label {cqz}
\end{equation}
with the electric form factor given by
\begin{equation}
{\tilde{G}}^2_{E}(Q^2) = \left [G^2_{Ep}(Q^2) + {\frac {N}{Z}}
G^2_{En}(Q^2) \right ] {\frac {(1 + \tau)} {(1 + 2\tau)}},
\end{equation}
where $Z$ and $N$ are number of protons and neutrons of the
target, respectively. $G_{Ep}$ and $G_{En}$ are the Sachs electric
form factors for the protons and neutrons, respectively. The last
factor corresponds to the relativistic correction factor, in which
$\tau$ is given by $\tau = Q^2 /4M_{N}^2$ with the nucleon mass
$M_{N}$. The lower limit $\omega_{min}$ in the integration
includes all inelastic contributions but excludes the elastic
peak.

In reviewing the Bates and Saclay datasets we do not find
experimental data at fixed $\omega$ and $q$ values within given
kinematics enough to obtain the longitudinal and transverse
structure functions from the Rosenbluth separation in Eq.
(\ref{stot}). Thus, expanding the range of the momentum transfer
at fixed $\omega$ value, we choose that the $q$ range at 350 MeV/c
is between 320 MeV/c and 380 MeV/c and at $q$=450 MeV/c from 420
MeV/c to 480 MeV/c to obtain proper separation. Furthermore,
Saclay group used the effective momentum approximation (EMA) to
include the electron Coulomb distortion but this has been proved a
poor approximation within the intermediate electron energy range
from 300 to 600 MeV although it is good at high electron energy
greater than 1.0 GeV \cite{kim96,kim01,kim05}.

\begin{figure}
\includegraphics[width=0.49\linewidth]{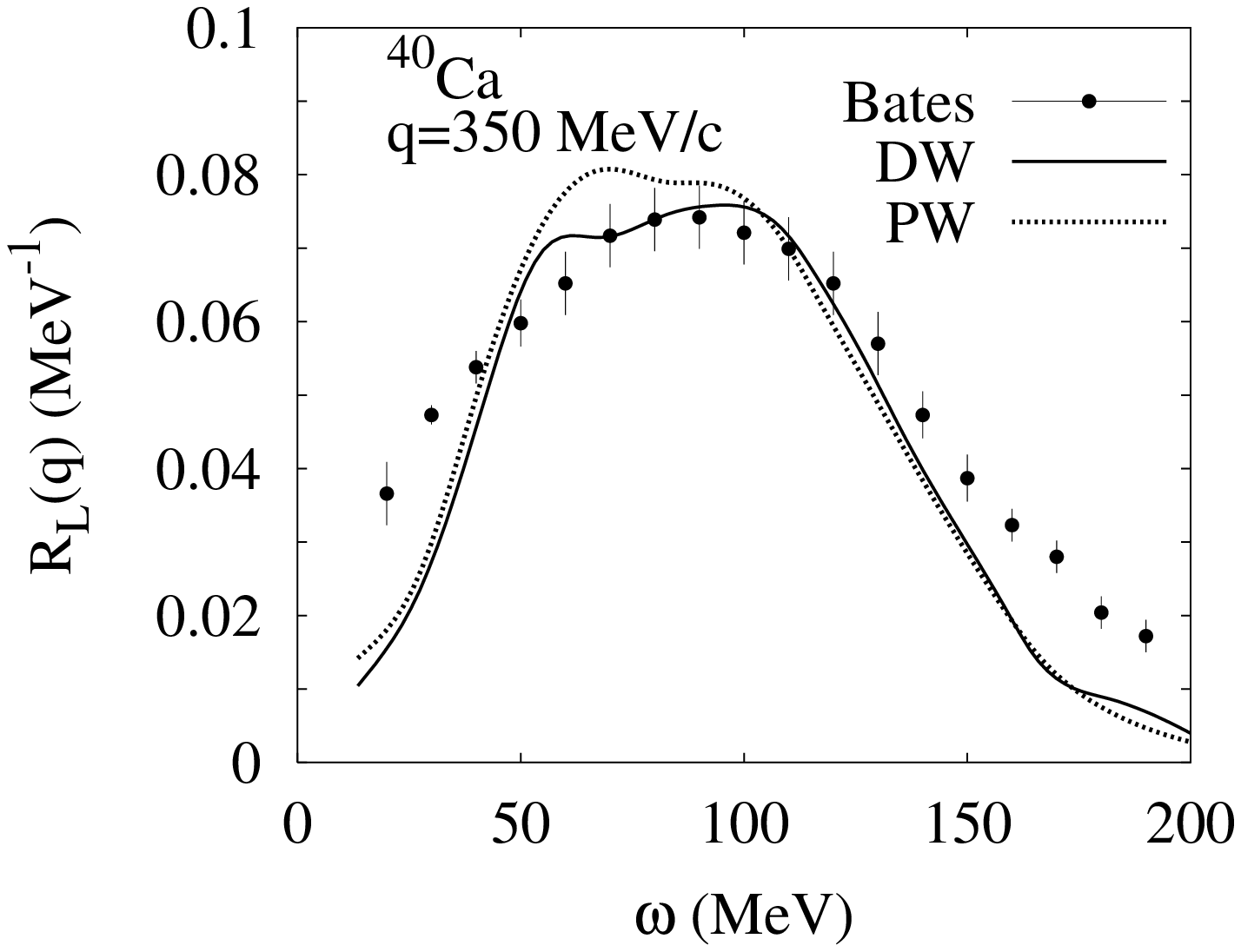}
\includegraphics[width=0.49\linewidth]{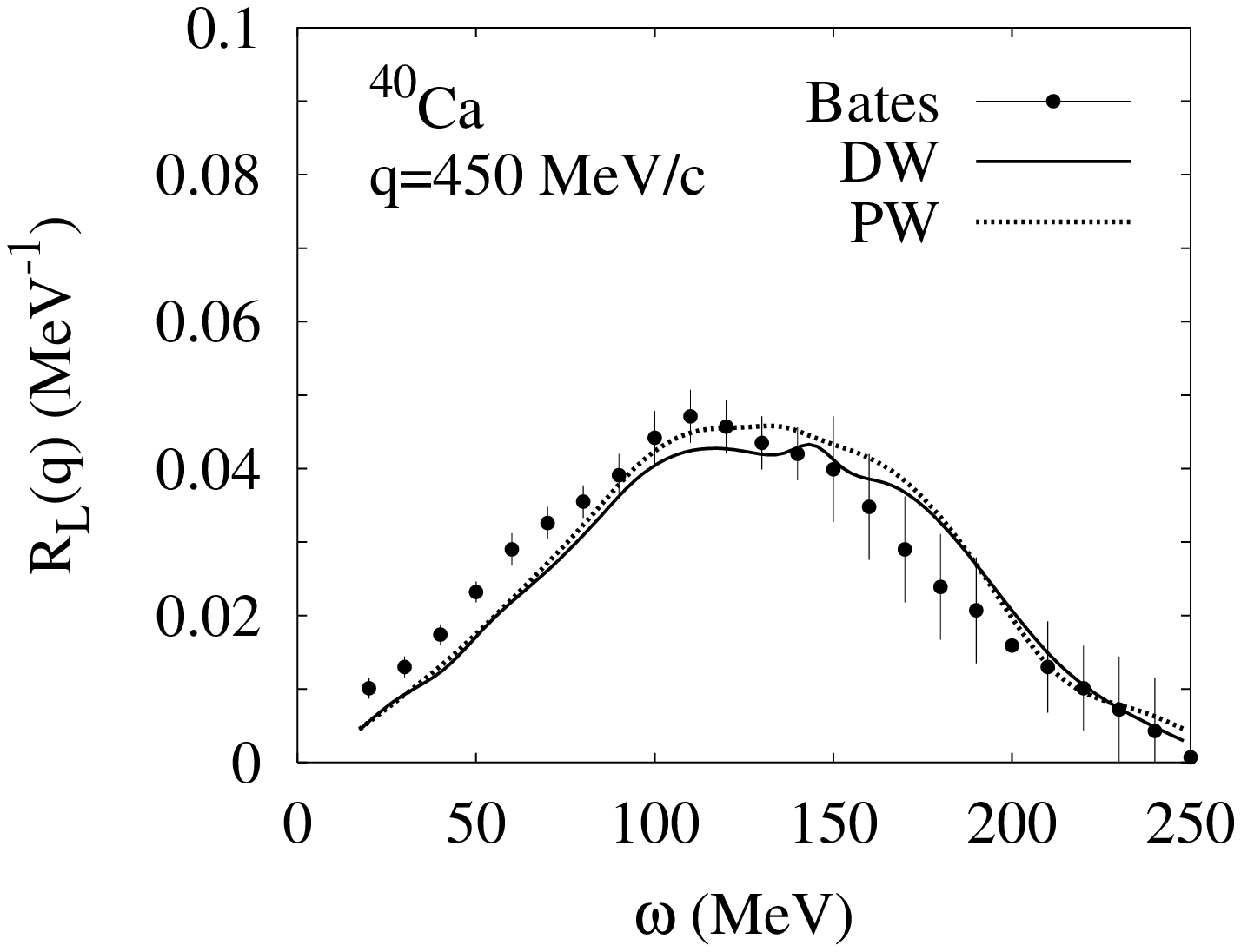}
\caption{The longitudinal structure functions of $^{40}$Ca at
$q$=350, 450 MeV/c. The solid lines are the results with the
inclusion of the electron Coulomb distortion and the dotted curves
are without the Coulomb distortion.}
\label{calq}\end{figure}

\begin{figure}
\includegraphics[width=0.49\linewidth]{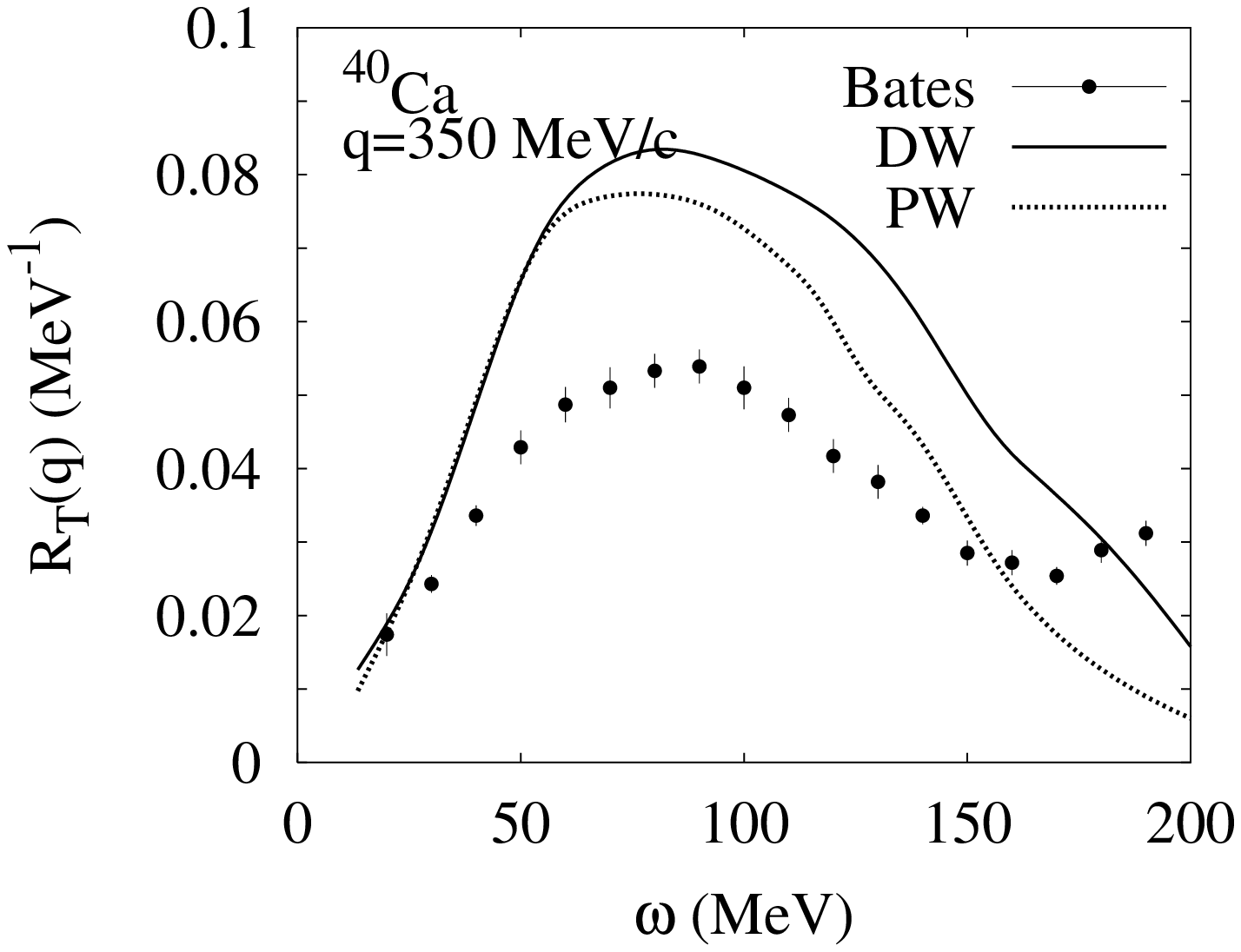}
\includegraphics[width=0.49\linewidth]{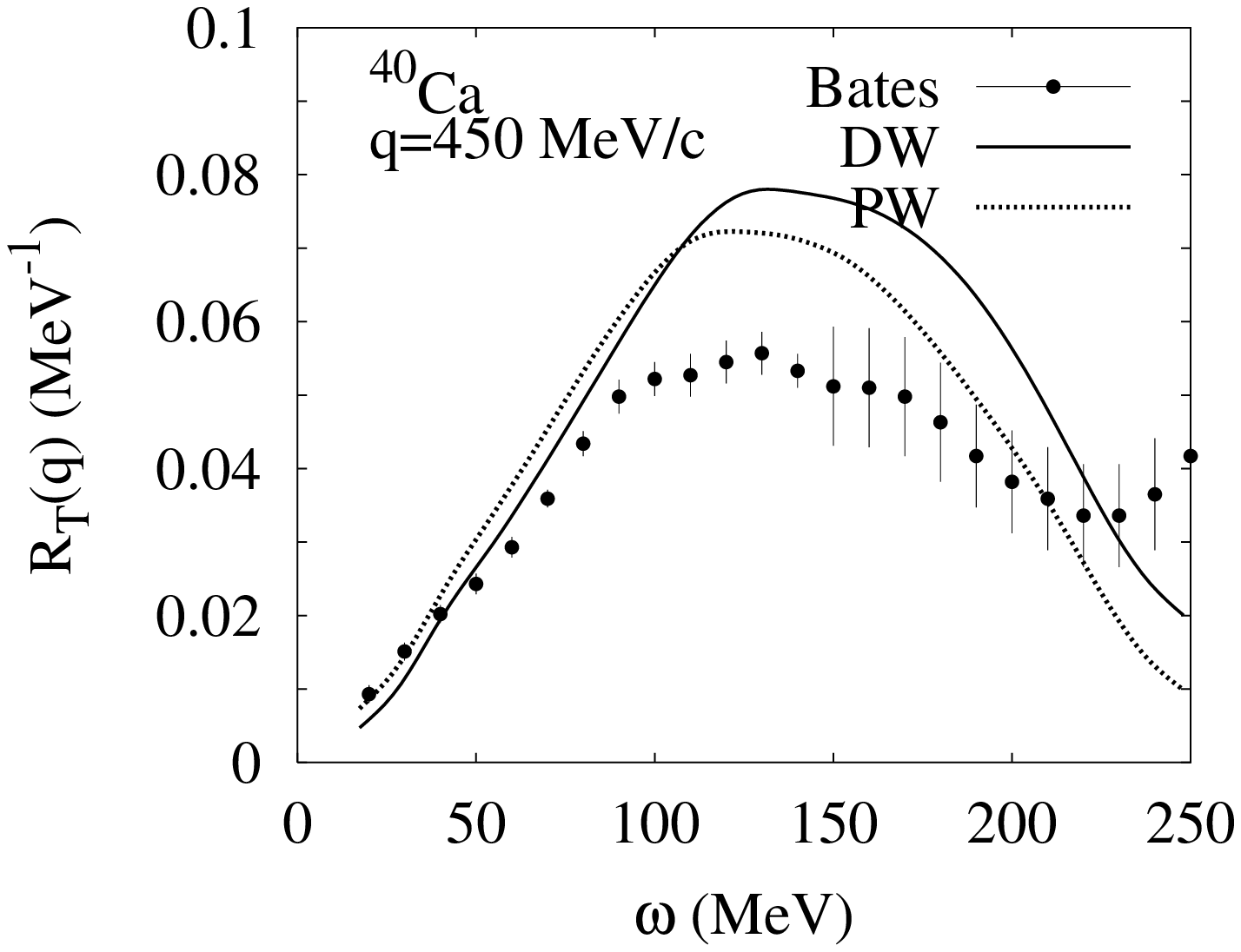}
\caption{The transverse structure functions of $^{40}$Ca at
$q$=350, 450 MeV/c. The solid lines are the results with the
inclusion of the electron Coulomb distortion and the dotted curves
are without the Coulomb distortion.}
\label{catq}\end{figure}

Figures \ref{calq} and \ref{catq} show the longitudinal and
transverse structure functions obtained from the slops and the
intercepts in Eq. (\ref{stot}) in terms of the energy transfer at
the momentum transfer $q$=350, 450 MeV/c on $^{40}$Ca. The solid
curves (labelled DW) represent the results with inclusion of the
electron Coulomb distortion in Ref. \cite{kim03}. The dotted
curves (labelled PW) are calculated without the electron Coulomb
distortion which uses the plane wave for the electrons. The
experimental data are taken from Bates \cite{bates}. Since the
final electron energy decreases with higher energy transfer, the
Coulomb effect in the transverse functions is larger than that in
the longitudinal functions in this region. In particular, in this
high energy transfer region, the short-range interactions lead to
the correlations between nucleons in the mean field. Hence the
effect of the Coulomb distortion affects the magnetization of
nucleons.

Our theoretical results show good description of the experimental
data in the longitudinal structure function but they do not in the
transverse functions in the magnitude although they have similar
shape. It should be noted that any other processes except the
quasieleastic scattering are not included in this calculation.

\begin{figure}
\includegraphics[width=0.49\linewidth]{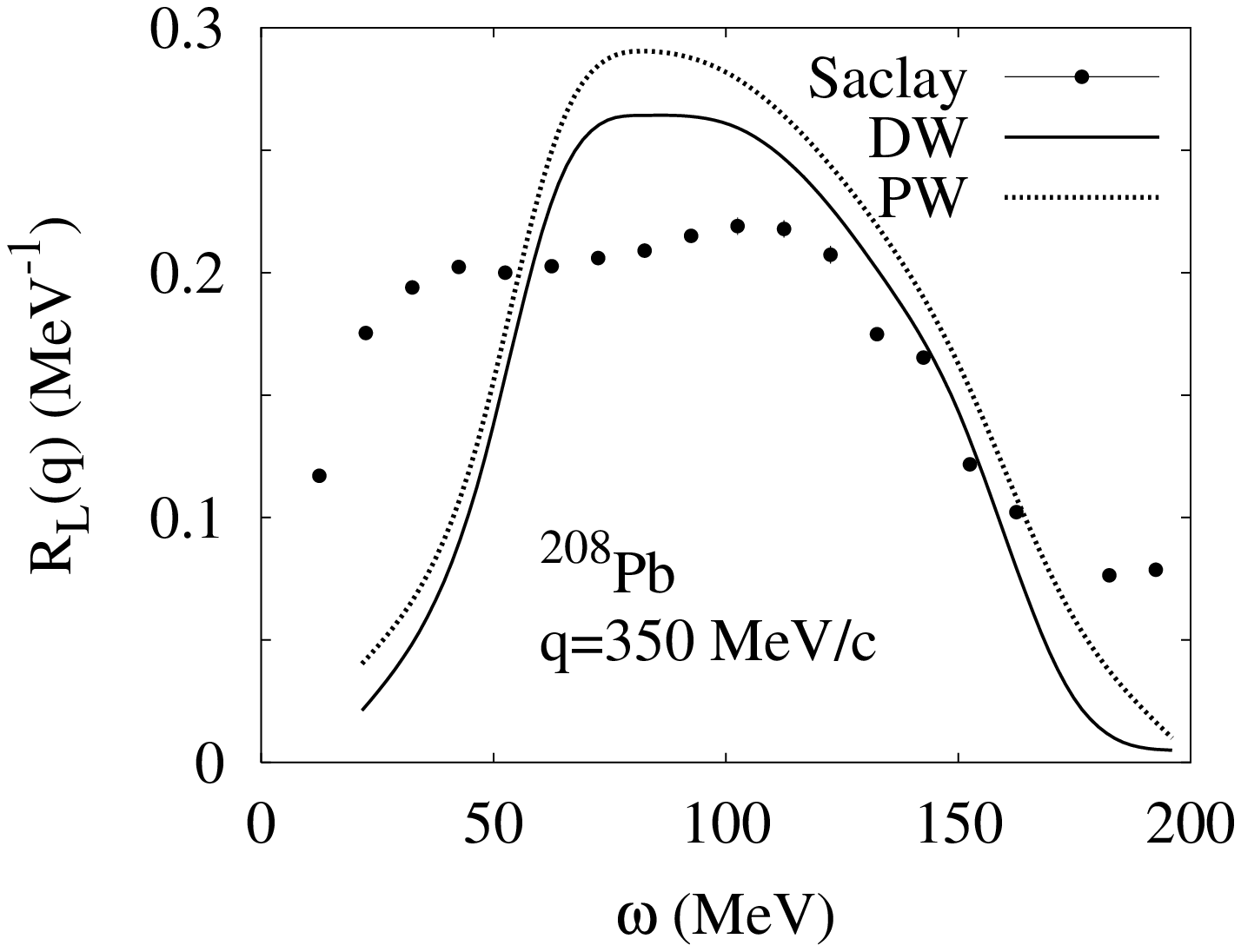}
\includegraphics[width=0.49\linewidth]{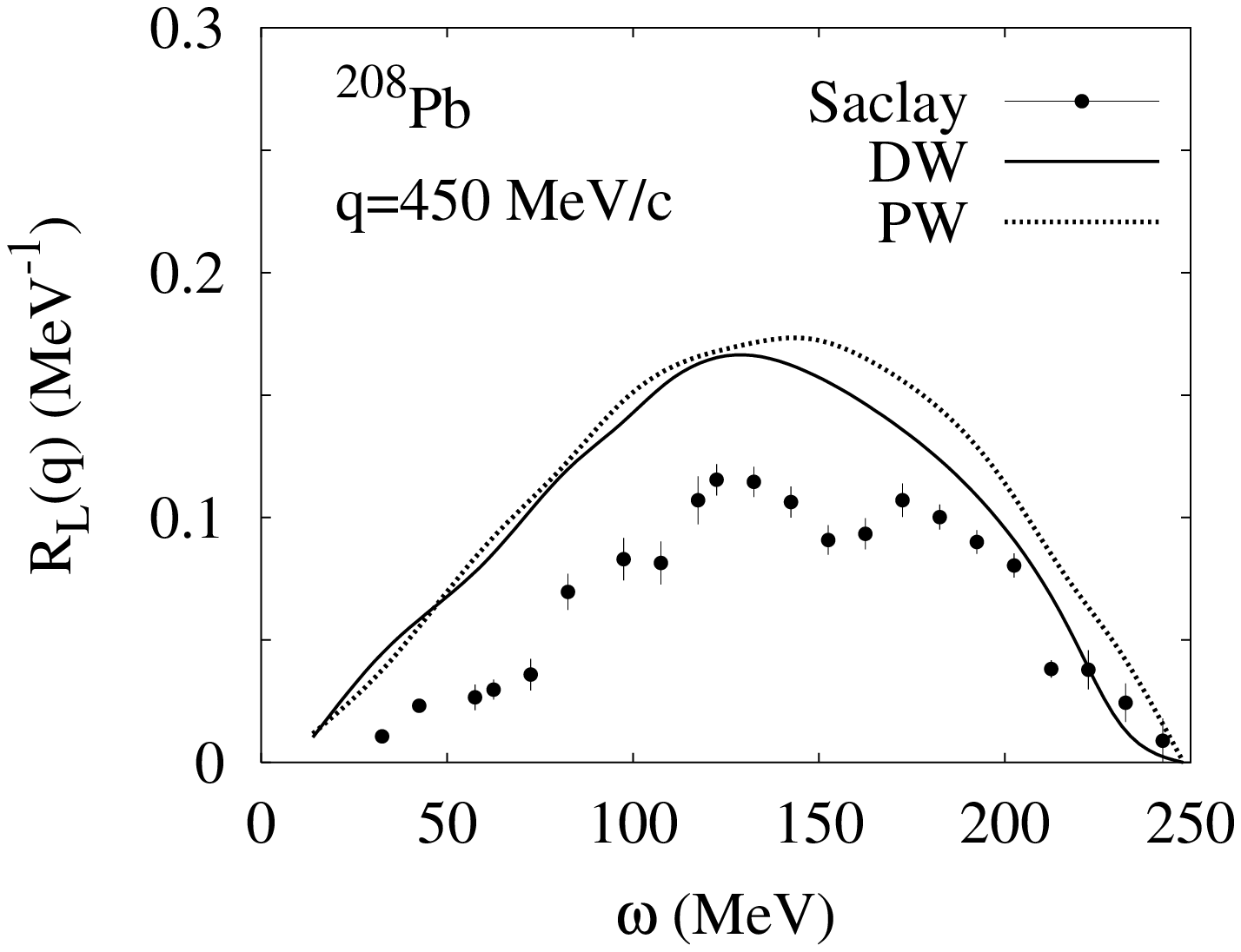}
\caption{The longitudinal structure functions of $^{208}$Pb at
$q$=350, 450 MeV/c. The solid lines are the results with the
inclusion of the electron Coulomb distortion and the dotted curves
are without the Coulomb distortion.}
\label{pblq}\end{figure}

\begin{figure}
\includegraphics[width=0.49\linewidth]{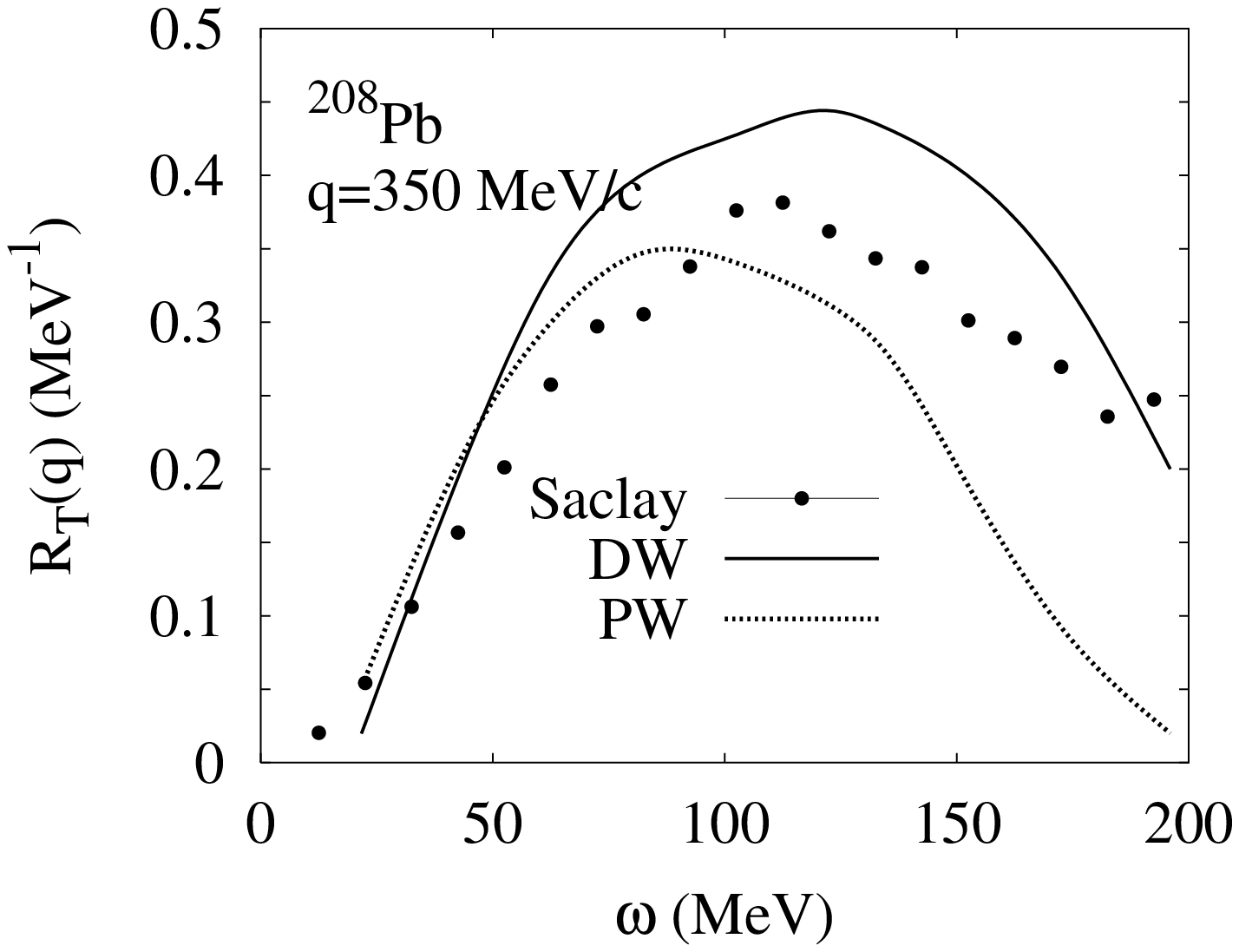}
\includegraphics[width=0.49\linewidth]{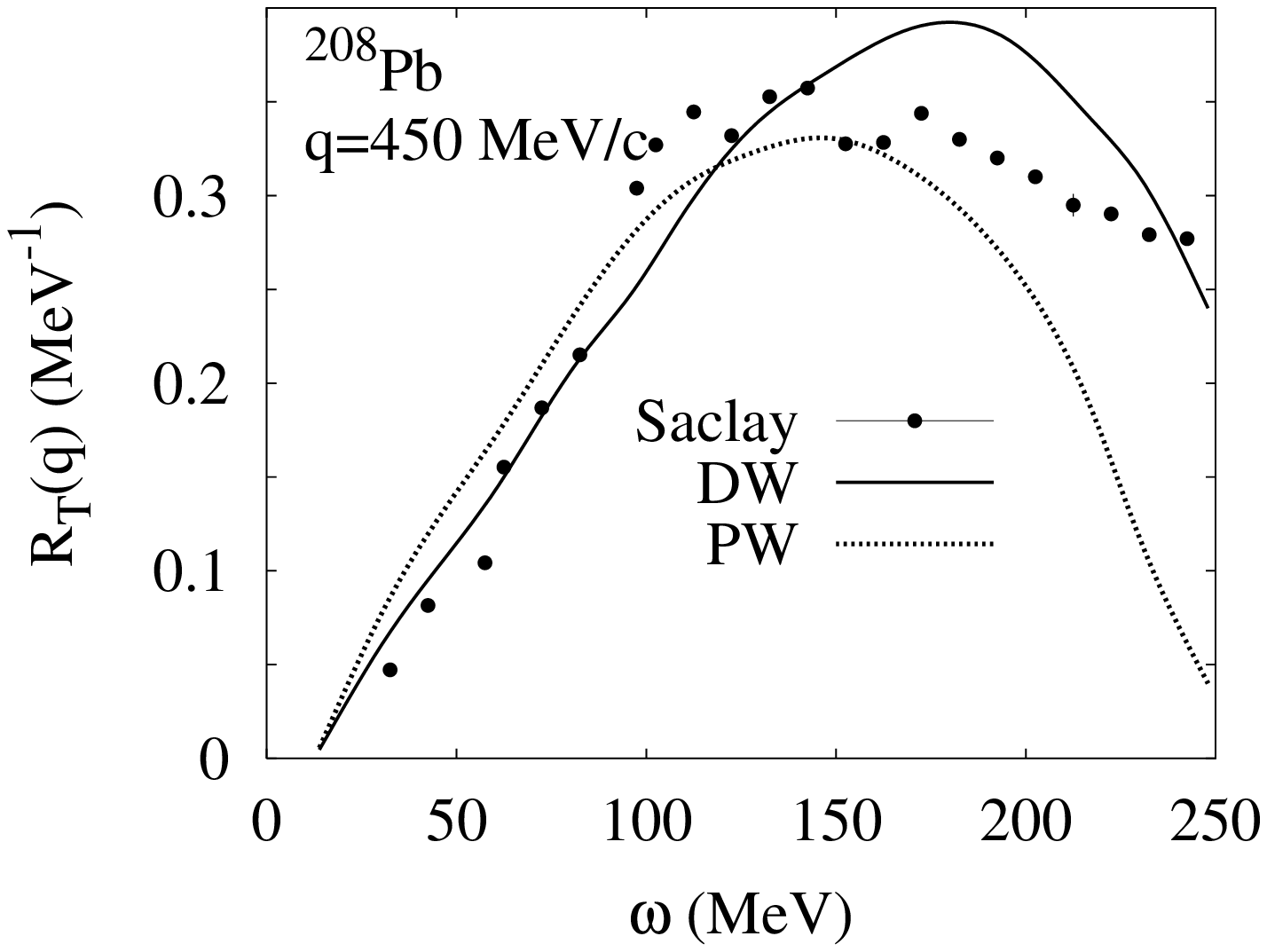}
\caption{The transverse structure functions of $^{208}$Pb at
$q$=350, 450 MeV/c. The solid lines are the results with the
inclusion of the electron Coulomb distortion and the dotted curves
are without the Coulomb distortion.}
\label{pbtq}\end{figure}

In Figs. \ref{pblq} and \ref{pbtq}, we compare our theoretical
calculations with Saclay experimental data \cite{sacl} of
$^{208}$Pb for the longitudinal and transverse structure functions
in terms of $\omega$ at the momentum transfer $q$=350, 450 MeV/c.
Like our previous works \cite{kim96,kim01}, our results are not
good descriptions of the longitudinal structure function with the
experimental data even in shape even if we take into account that
the Saclay data were given in terms of the effective momentum.
However, they relatively well produce the transverse parts.

\begin{figure}
\includegraphics[width=0.49\linewidth]{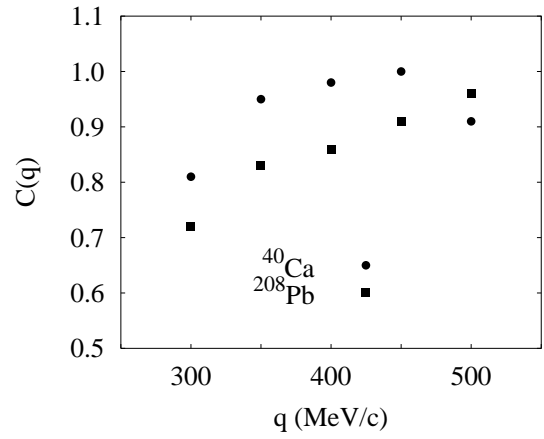}
\caption{The Coulomb sum rule for our model in terms of $q$
values. The solid circles are for $^{40}$Ca and the solid
rectangles are for $^{208}$Pb, respectively.} \label{csrq}
\end{figure}

In order to calculate the CSR we need to know the lower and upper
limits of the integration in Eq. (\ref{cqz}). While the different
values were used for the lower limit $\omega_{min}$ with each
momentum transfer in Ref. \cite{sacl}, we choose $\omega_{min}=10$
MeV which is enough to exclude the elastic process in all
calculations. For the upper limit $\omega_{max}$ of the
integration we use different values for each momentum transfer,
$\omega_{max}=200$ MeV for $q$=300 and 350 MeV/c,
$\omega_{max}=225$ MeV for $q$=400 MeV/c, and $\omega_{max}=250$
MeV for $q$=450 and 500 MeV/c. In these upper limits we obtain the
Coulomb sum rule $C(q)$ values within our theoretical model in
Fig. \ref{csrq}. Our calculations of $C(q)$ for $^{40}$Ca show
tendency similar to the Bates data \cite{bates} but those for
$^{208}$Pb do not match with the Saclay data \cite{sacl,morg}.
From these results, there is no suppression of the longitudinal
structure functions so much as 50 \% obtained from Saclay group
\cite{sacl}. Our results exhibit the nucleus-dependence of the
CSR.

In this report, we examined the suppression of the longitudinal
structure function in the CSR within the relativistic single
particle model for the inclusive $(e,e')$ reaction in the
quasielastic region. Our theoretical calculations show a good
description of the experimental data for the longitudinal
structure function of $^{40}$Ca but do not have good expressions
for the transverse part on $^{40}$Ca and for both functions on
$^{208}$Pb. The electron Coulomb distortion in the transverse part
appears larger than that in the longitudinal part with higher
energy transfer. Our model shows the nucleus-dependence of the
CSR. We find that the suppression (about 50\%) of the longitudinal
structure function within our theoretical model is not so much as
Saclay group insisted. Nevertheless our results show some
deviation of the Coulomb sum rule from unity.

\newpage

\end{document}